\documentclass[aps,superscriptaddress,nofootinbib,floatfix]{revtex4-2}

\usepackage{amsmath}
\usepackage{mathrsfs}
\usepackage{amssymb}
\usepackage{bm} 
\usepackage{graphicx,grffile,textcomp}
\usepackage{dcolumn} 
\usepackage[usenames,dvipsnames]{xcolor}
\usepackage{multirow}
\usepackage{epstopdf}
\usepackage{mathtools} 
\usepackage{tabularx}
\usepackage{microtype}
\usepackage{comment}
\usepackage[version=4]{mhchem} 
\usepackage{array}
\usepackage{float}
\usepackage{wrapfig} 
\usepackage{url}
\usepackage{hyperref}
\usepackage{cleveref}
\usepackage[utf8]{inputenc}
\usepackage[T1]{fontenc}
\usepackage{bbold}

\hypersetup{colorlinks=true, linkcolor=blue, urlcolor=blue, citecolor=blue}

\definecolor{myred}{RGB}{230, 7, 77}

\definecolor{blue(ryb)}{rgb}{0.01, 0.28, 1.0}
\definecolor{jade}{rgb}{0.0, 0.66, 0.42}
\definecolor{cadmiumred}{rgb}{0.89, 0.0, 0.13}
\definecolor{darkviolet}{rgb}{0.58, 0.0, 0.83}

\begin{document}

\title{Translocation of a daughter vesicle in a model system of self-reproducing vesicles}

\author{Manit Klawtanong}
\affiliation{Department of Physics, Ramkhamhaeng University, Bang Kapi, Bangkok 10240, Thailand} 

\author{Yuka Sakuma}
\affiliation{Department of Physics, Tohoku University, Aoba, Aramaki, Aoba-ku, Sendai 980-8578, Japan}

\author{Masayuki Imai}
\affiliation{Department of Physics, Tohoku University, Aoba, Aramaki, Aoba-ku, Sendai 980-8578, Japan}

\author{Toshihiro Kawakatsu}
\affiliation{Department of Physics, Tohoku University, Aoba, Aramaki, Aoba-ku, Sendai 980-8578, Japan}

\author{Petch Khunpetch}
\email{petch.k@rumail.ru.ac.th}
\thanks{Corresponding author}
\affiliation{Department of Physics, Ramkhamhaeng University, Bang Kapi, Bangkok 10240, Thailand}

\begin{abstract}
Translocation of a daughter vesicle from a mother vesicle through a pore is experimentally studied by many groups using a model system of self-reproducing vesicles. However, the theoretical formulation of the problem is not fully understood. In the present study, we present a theoretical formulation of the process based on our previous work [P. Khunpetch et al., Phys. Fluids \textbf{33}, 077103 (2021)]. In our previous work, we considered the daughter vesicle as a rigid body. In the present work, however, we allow the daughter vesicle to deform during the expulsion process. We thus derive the free energy constituting of the elastic moduli of both the mother and daughter vesicles, and of pressure-driven contribution. The minimum energy path of the translocation is searched by using the string method. Our improved model successfully suggests the disappearance of the energy barrier where all the elastic moduli are in agreement with the experimental reports, while the previous work is unsuccessful to do so.
The equations of motion of the daughter vesicle have been derived within the framework of the Onsager principle. We found that the translocation time of the daughter vesicle can be reduced when the pressure inside the mother vesicle increases, or the initial size of the daughter vesicle decreases.
\end{abstract}

\maketitle

\section{Introduction}
 \label{sec:intro}
 
The formation of amphiphilic vesicles is an important step of the pathways toward protocells that perform the minimal functions for cellular life, which provides an insight into the origin of life.~\cite{SI1, Small, Chang} A vesicle can be regarded as a container for a protocell that can show self-reproduction incorporating with converting the energy from its environment to support the activity (metabolism).~\cite{L, RBCDKPS, DS, DD}
The model systems of self-reproducing vesicles are reported by many authors.~\cite{HFS, TTS, TS, WWL, SI} Typically, there are
two topological pathways of formation of daughter vesicles: (1) \textbf{Translocation.} Smaller vesicle(s) is (are) formed inside the mother vesicle and then be expelled through a transient pore on the mother vesicle. (2) \textbf{Budding.} The mother vesicle can deform to a pear shape and splits into two vesicles. 


The translocation of a daughter vesicle through a pore is demonstrated by many groups.~\cite{WWL, SI, KSGBPW} Previous studies have shown that an original giant unilamellar vesicle (GUV) can turn its molecules to assemble a daughter vesicle with or without the help of catalyst.~\cite{SI} As the daughter vesicle grows to a certain radius, the pore on the mother vesicle starts to open. For a small daughter vesicle, the leaking fluid will drag the daughter toward the pore and, finally, the daughter can exit the mother before the pore closes.~\cite{KSGBPW} 
The situation is rather different if the daughter vesicle is larger than the size of the pore. In such a case, the daughter vesicle is dragged by the leaking fluid toward and plugs the pore. The daughter vesicle can be stuck at this stage for a few minutes. The pressure inside the mother vesicle increases because the membrane is building up the tension. (We might assume that there is no liquid leaking out from the mother vesicle. Thus, the pressure pushing the daughter vesicle out does not drop.) Due to its flexibility, the daughter can squeeze itself trying to pass through the pore. The translocation can be successful or the daughter vesicle can be cut by closing pore.~\cite{YS, KSGBPW} 

\begin{figure}[]
\includegraphics[scale=0.4]{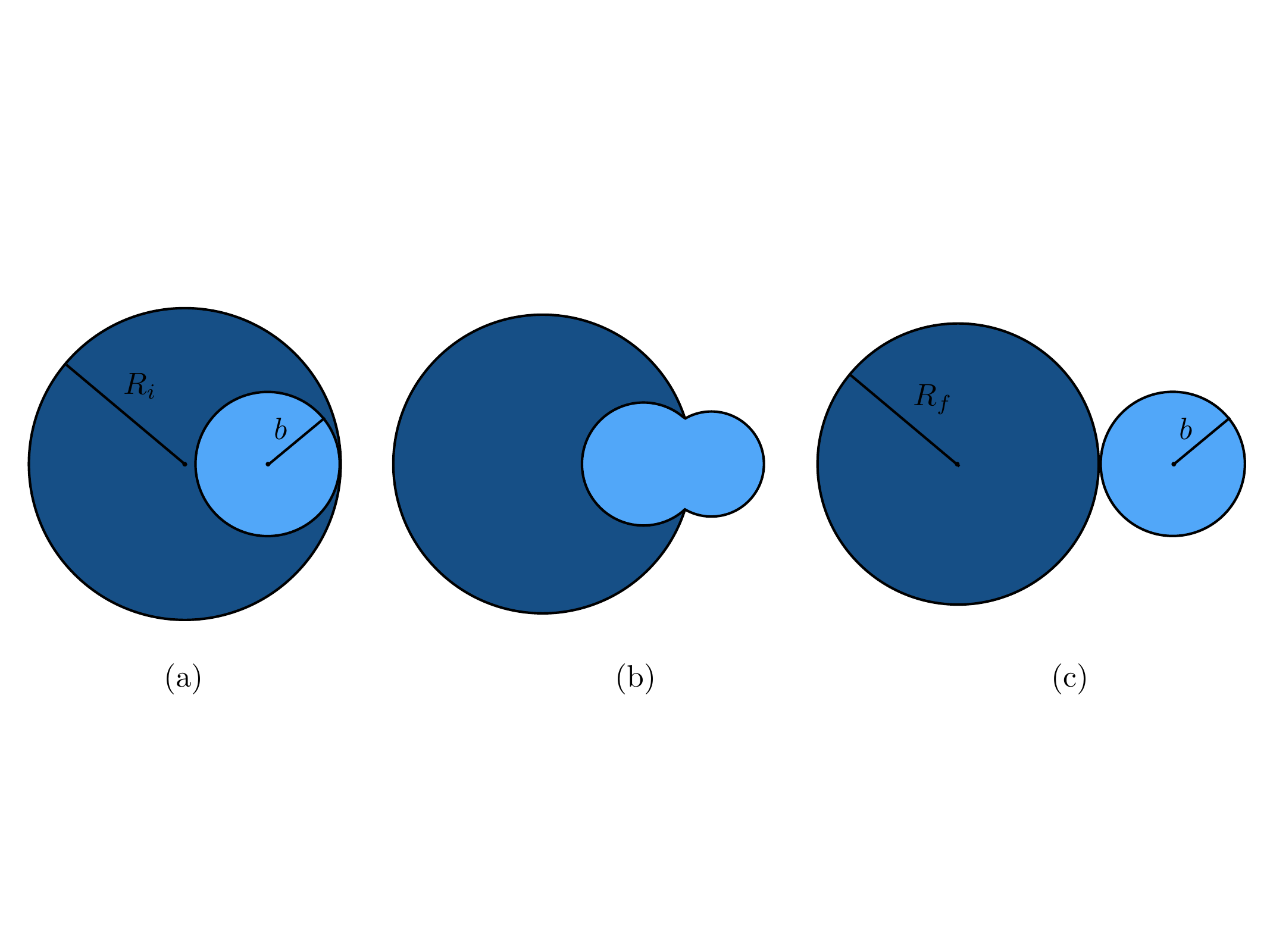}%
\caption{\label{states}Three states of the daughter vesicle's translocation. (a) The initial state. The daughter vesicle with radius $b$ is immersed in the liquid inside the mother. (b) The intermediate state. The daughter vesicle is expelled through the pore on the surface of the mother vesicle which gradually expands. (c) The final state. The daughter vesicle is completely discharged from the mother vesicle. The size of the mother vesicle shrinks $(R_f < R_i)$. The pore is absolutely sealed.}
\end{figure}


Theoretical study done by Khunpetch et al.~\cite{KSIK} has shown that, regarding the daughter vesicle as a spherical rigid body, the daughter can be discharged due to the surface tension of the mother vesicle. However, the model is unable to predict the process with the elastic moduli in accordance with the reported experiments. In this present study, we have developed a new model based on our previous work.~\cite{KSIK} We have considered a deformable daughter vesicle when it passes through the pore on the mother vesicle. Due to pressure difference between the inside and outside regions of the mother vesicle, the daughter vesicle is then expelled through the pore. Based upon the framework of Helfrich theory for lipid bilayers~\cite{H} widely used for studying the problems of lipid bilayer vesicles, for example, the encapsulation of a rigid particle by a vesicle,~\cite{MM} translocation of a vesicle through a pore,~\cite{SM, KMKD} and simulation of the overdamped dynamics of membranes and vesicles,~\cite{Wolgemuth} we have searched the minimum energy path (MEP) of the translocation by using the string method.~\cite{ERV1, Ren, ERV2} In the present study, we will show that our improved model successfully suggests the disappearance of the energy barrier where all the elastic moduli are in agreement with the experimental reports, while the previous work is unsuccessful to do so.

We have employed the Onsager variational principle proposed by L. Onsager~\cite{O1, O2} to study the dynamics of the system. In his seminal works, Onsager derived his reciprocal relations for a nonequilibrium thermodynamic system involving irreversible processes from 
the general theory of fluctuations and the principle of microscopic reversibility. Moreover, his variational principle is considered as an  extension of Rayleigh’s principle of the least dissipation of energy.~\cite{D2} In recent years, Onsager principle has been widely used for the derivation of the evolution equations in soft matter dynamics.~\cite{D0, Doi2021, MD, Oya} The principle states that if the system is characterized by a set of variables $\mathbf{q} = (q_1, q_2, ..., q_f)$, the evolution equation can be derived by minimizing the Rayleighian of the system defined by
\begin{equation}  
\mathcal{R}(\mathbf{\dot{q}}) = \Phi (\mathbf{\dot{q}}) + \dot{F}(\mathbf{q}),
\end{equation}
with respect to $\mathbf{\dot{q}}$, where $\Phi$ is the dissipation function which is quadratic in $\mathbf{\dot{q}}$ and $\dot{F}$ is the time derivative of the free energy in the isothermal system. Within this framework, we will derive the evolution equations of the daughter vesicle and we will show that the translocation time of the daughter vesicle decreases as the size of the daughter decreases, or the pressure inside the mother vesicle increases.

We organize the paper as follows. In Sec.~\ref{sec:model}, we describe our improved model. Section~\ref{sec:results and discussions} is devoted to the discussions of the results. Finally, we conclude this study in Sec.~\ref{sec:conclusions}.

\section{Theory}
 \label{sec:model}
 
 \subsection{\label{subsec:Geometry}Geometry of the system}

\begin{figure}[]
\includegraphics[scale=0.4]{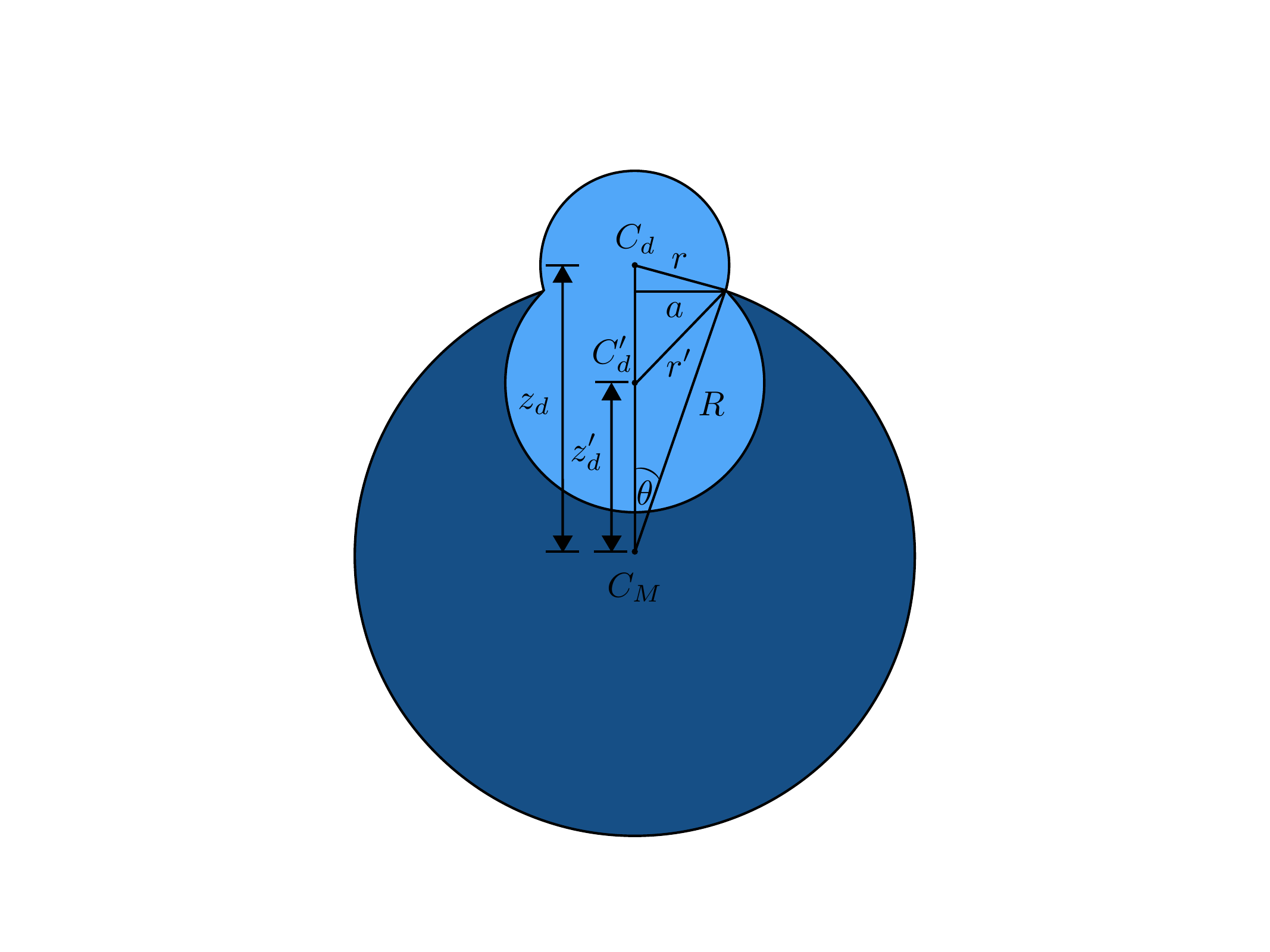}%
\caption{\label{model}The intermediate state of the translocation of the deformed daughter vesicle. $r$ and $r'$ are the radii of the spherical caps indicating the transmitted and inclusion parts of the daughter vesicle, respectively. The center of the mother vesicle is denoted as $C_M$. The centers of the spheres for the transmitted and inclusion parts of the daughter vesicle are $C_d$ and $C_{d}'$, respectively. $z_d$ and $z_{d}'$ are the distances from $C_M$ to $C_d$ and to $C_{d}'$, respectively. $R$ is the radius of the mother vesicle and $a$ is the radius of the pore, both of which changes with time as the daughter vesicle translocates from the mother vesicle. Dark blue region shows the liquid volume between the mother and daughter vesicles.}
\end{figure}

We consider the translocation of a deformable daughter vesicle through a pore of a mother vesicle as a model system of self-reproduction vesicles, as shown in Fig.~\ref{model}.
In our consideration, the thickness of the membrane of the daughter and mother vesicles is neglected. The approach described below is broadly based on our previous work~\cite{KSIK}. Here we repeat the model and emphasize the modifications we have done in this work. In the improved model, the daughter vesicle is modelled as
a set of two connected spherical caps, one is inside the mother vesicle (remaining part) and the other is outside it (discharged part). The discharged part from the pore is defined by the spherical cap with radius $r$ cut off by a plane of the pore and the remaining part inside the mother vesicle yet is defined as a part of a sphere with radius $r'$. We denote the center of the mother vesicle as $C_{M}$ and the centers of the spheres for discharged and remaining parts of the daughter vesicle as $C_{d}$ and $C_{d}'$, respectively. The distances from $C_{M}$ to $C_{d}'$ and to $C_{d}$ are $z_{d}'$ and $z_{d}$, respectively. During the translocation, the pore gradually enlarges and then reseals due to the line tension energy. The pore is assumed to be a circle  with radius $a$ where its center is located on the line connecting the center of the mother vesicle and the centers of the two parts of the daughter vesicle. The angle $\theta$ is defined by the line joining $C_{M}$, $C_{d}'$ and $C_{d}$, and the line between $C_{M}$ and the pore's rim. The radius of the mother vesicle is denoted as $R$, which is a function of time as the daughter vesicle translocates from the mother vesicle. We have assumed that the contact line between the pore's perimeter and the deformed daughter vesicle is so tight that liquid did not flow out of the mother and there is no liquid layer between the outer surface of the daughter and the inner edge of the pore. 
In our assumption, the daughter vesicle with initial radius $b$ is expelled due to 
the pressure difference between inside and outside of the mother vesicle. In the initial state, the mother vesicle has radius $R_{0}$ and the daughter vesicle is slightly discharged from the mother with $r = r_0$ and $\theta =\theta_0$. 
When the daughter vesicle entirely passes through the pore of the mother vesicle, we would have $r=b$ and $\theta =0$. The values of the parameters $z_{d}$ and $z_{d}'$ at the initial states are denoted by $z_{d,0}$ and $z_{d,0}'$. From our model, the variables $z_{d}$, $z_{d}'$, $r$, $r'$, $R$ and $\theta$ are geometrically constrained through
\begin{equation}
\label{cosine law}
\cos\theta=\frac{R^{2}+z_{d}^{2}-r^{2}}{2z_{d}R}=\frac{R^{2}+z_{d}'^{2}-r'^{2}}{2z_{d}'R},
\end{equation}
and the volumes of the mother and daughter vesicles are 
\begin{eqnarray}
\label{volume mother}
V_{M}&=&\frac{4}{3}\pi R^{3}-\pi r'^{2}\Big[(R\cos\theta-z_{d}')+r'\Big]+\frac{\pi}{3}\Big[(R\cos\theta-z_{d}')^{3}+r'^{3}\Big]\nonumber\\
&&-\pi R^{3}(1-\cos\theta)+\frac{\pi R^{3}}{3}(1-\cos^{3}\theta),
\end{eqnarray}
and

\begin{eqnarray}
\label{volume daughter}
V_{d}&=&\pi r'^{2}\Big[(R\cos\theta-z_{d}')+r'\Big]-\frac{\pi}{3}\Big[(R\cos\theta-z_{d}')^{3}+r'^{3}\Big]\nonumber\\
&&+\pi r^{2}(z_{d}+r-R\cos\theta)-\frac{\pi}{3}\Big[r^{3}-(R\cos\theta-z_{d})^{3}\Big],
\end{eqnarray}
respectively.
Since the volumes of the two vesicles are preserved, this leads us to $\Delta V_M = \Delta V_d = 0$. Using the specified constraints, $z_{d}'$, $r'$, $R$, and $\theta$ can be eliminated such that the free energy is a function of $r$ and $z_{d}$ only. (Details of the eliminations can be found in SI.)



\subsection{\label{subsec:free energy}The free energy}

At equilibrium, the daughter vesicle is located inside the mother vesicle where the surface areas of the daughter and mother vesicles are $4\pi r_\textrm{eq}^2$ and $4\pi R_{\textrm{eq}}^2$, respectively. The total free energy of the system is
\begin{equation}
\label{Ftot}
  F = F_{M} + F_{d} + P(V_{0}-V_{\textrm{tran}}), 
\end{equation}
where $F_M$ and $F_d$ are the free energies of the mother and daughter vesicles, respectively.
The last term in Eq.~\ref{Ftot} is pressure-driven contribution to the free energy where $P$ is the pressure inside the mother vesicle. (We have set the pressure outside the mother vesicle as zero without loss of generality, since only the pressure difference between the inner and outer regions is meaningful.) The initial volume of the daughter vesicle is $V_0 = 4\pi b^3/3$. $V_{\textrm{tran}}$ is the partial volume of the daughter vesicle that is already expelled from the mother vesicle. 

\subsubsection{\label{subsubsec:FM}The mother vesicle's free energy $F_M$}

The free energy of the mother vesicle is given by
\begin{equation}
\label{FM1}
 F_{M}=F_{M}^{\textrm{bending}} +
F_{M}^{\textrm{stretching}} +
F_{M}^{\textrm{line}}.
\end{equation}
The first term on the right hand side of Eq.~\ref{FM1} is the Helfrich bending energy~\cite{H} which can be written as
\begin{equation}
\label{FM2}
F_{M}^{\textrm{bending}}=\Bigg[\frac{\kappa}{2}\int(2H_{M}-c_{0})^{2}dA_{M}+\kappa_{G}\int KdA_{M}\Bigg],
\end{equation}
where $H_M$ is the mean curvature of the mother vesicle and it is defined as $H_M = (1/2)(1/R_1 +1/R_2)$ ($R_1$ and $R_2$ are the radii of principal curvatures.). For the spherical mother vesicle with radius $R$, $H_M=1/R$. $c_{0}$ is the spontaneous curvature which is zero for symmetric bilayers. Because the topology of the mother vesicle has not been preserved, we must take the Gaussian curvature, $K$, into account, which is defined as $K = 1/(R_1 R_2)$ and can be simply $1/R^2$ for the spherical mother vesicle. The local bending rigidity $\kappa$ is of the order of $10\, k_B T$ which was experimentally obtained in~\cite{ROMNE}. The Gaussian bending rigidity $\kappa_{G}$ has a negative sign indicating that the surface with lower genus is preferred. The measurement shows that, for example, $-\kappa_G/\kappa = 0.9\pm0.38$ for the various mixing ratios of [dioleoyl-\textit{sn}-glycero-3-phosphatidylcholine (DOPC)]/[sphingomyelin (SM)]/[cholesterol (Chol)].~\cite{BDWJ} The ratio $(-\kappa_{G}/\kappa)$ for bilayers can also be found in~\cite{HBD}. In the present study, we will consider for isotropic liquid membranes with vanishing spontaneous curvature $(c_0 = 0)$ where $\kappa_{G}/\kappa=-1$ (see Appendix A for details of the proof.).

The integral $\int dA_M$ is clearly the area of the spherical mother vesicle subtracted by the area of the pore which can be read
\begin{equation}
\label{surface area}
A_M = 2\pi R^{2}(1+\cos\theta).
\end{equation}
Thus, the Helfrich bending energy for the mother vesicle is
\begin{equation}
\label{bending_Mother}
F_{M}^{\textrm{bending}} = \pi\kappa(1+\cos\theta)\Bigg(R^{2}\Bigg(\frac{2}{R}-c_{0}\Bigg)^{2}+\Bigg(\frac{2\kappa_{G}}{\kappa}\Bigg)\Bigg).
\end{equation}
By using the geometrical constraint, $\cos\theta=(R^{2}+z_{d}^{2}-r^{2})/(2z_{d}R)$, (Eq.~(\ref{cosine law})), the bending energy becomes
\begin{equation}
F_{M}^{\textrm{bending}}=\pi\kappa\Bigg(1+\frac{R^{2}+z_{d}^{2}-r^{2}}{2z_{d}R}\Bigg)\Bigg(R^{2}\Bigg(\frac{2}{R}-c_{0}\Bigg)^{2}+\Bigg(\frac{2\kappa_{G}}{\kappa}\Bigg)\Bigg).
\end{equation}


The second term on the right hand side of Eq.~(\ref{FM1}) is the stretching energy of the mother vesicle which can be written as
\begin{eqnarray}
\label{stretching_mother}
F_{M}^{\textrm{stretching}}&=&\frac{\lambda}{2}\frac{(A_M - A_{M, \textrm{eq}})^{2}}{A_{M, \textrm{eq}}}\nonumber\\
&=&\frac{\pi\lambda}{2R_{\textrm{eq}}^{2}}[R^{2}(1+\cos\theta)-2R_{\textrm{eq}}^{2}]^{2}\nonumber\\
&=&\frac{\pi\lambda}{2R_{\textrm{eq}}^{2}}\Bigg[R^{2}\Bigg(1+\frac{R^{2}+z_{d}^{2}-r^{2}}{2z_{d}R}\Bigg)-2R_{\textrm{eq}}^{2}\Bigg]^{2},
\end{eqnarray}
where, the constraint used for eliminating $\theta$ in the bending energy comes to the end for the same purpose. The value of the stretching modulus $\lambda$ can be found experimentally to be of the order of $10^{8}\, k_{B}T/\mu\textrm{m}^{2}$.~\cite{ER}


The last contribution is the line tension energy which could be given by
\begin{eqnarray}
\label{line tension_mother}
F_{M}^{\textrm{line}}&=&\sigma\oint dl\nonumber\\
&=&2\pi\sigma R\sin\theta.
\end{eqnarray}
Using $\cos\theta=(R^{2}+z_{d}^{2}-r^{2})/(2z_{d}R)$ with the identity $\sin^{2}\theta+\cos^{2}\theta=1$, the line tension energy is
\begin{equation}
F_{M}^{\textrm{line}}=2\pi\sigma R\sqrt{1-\frac{(R^{2}+z_{d}^{2}-r^{2})^{2}}{(2z_{d}R)^{2}}}.
\end{equation}
The reported line tension modulus $\sigma$ is of the order of $10^{3}-10^{4}\, k_{B}T/\mu\textrm{m}$ for bare DOPC bilayers.~\cite{KSGBPW}

Thus, the free energy of the mother vesicle which is a function of the variables $r$, $z_d$, and $R$ is given by
\begin{eqnarray}
\label{FM2}
F_{M}(r,z_{d},R)&=&\pi\kappa\Bigg(1+\frac{R^{2}+z_{d}^{2}-r^{2}}{2z_{d}R}\Bigg)\nonumber\\
& &\times\Bigg(R^{2}\Bigg(\frac{2}{R}-c_{0}\Bigg)^{2}+\Bigg(\frac{2\kappa_{G}}{\kappa}\Bigg)\Bigg)\nonumber\\
& &+\frac{\pi\lambda}{2R_{\textrm{eq}}^{2}}\Bigg[R^{2}\Bigg(1+\frac{R^{2}+z_{d}^{2}-r^{2}}{2z_{d}R}\Bigg)-2R_{\textrm{eq}}^{2}\Bigg]^{2}\nonumber\\
& &+2\pi\sigma R\sqrt{1-\frac{(R^{2}+z_{d}^{2}-r^{2})^{2}}{(2z_{d}R)^{2}}}.
\end{eqnarray}
By using the method to eliminate $R$ described in the SI, we will obtain $F_M = F_M [r, z_d]$.


\subsubsection{\label{subsubsec:FM}The daughter vesicle's free energy $F_d$}

We have assumed that the daughter vesicle has a spherical shape at equilibrium. The free energy of the daughter vesicle is given by
\begin{equation}
\label{Fd1}
 F_{d}=F_{d}^{\textrm{bending}} +
F_{d}^{\textrm{stretching}}.
\end{equation}
The bending energy of the daughter vesicle can be read
\begin{equation}
\label{bending daughter}
F_{d}^{\textrm{bending}}=\frac{\kappa}{2}\int(2H_{d}-c_{0})^{2}dA_{d},
\end{equation}
where $H_d$ is the mean curvature which is different for the two portions of the daughter vesicle, \textit{i.e.},
\begin{equation}
    H_{d}^{+} = \frac{1}{r}\, \, \, \,(\textrm{discharged part}),\, \, \, \, \, \, \, H_{d}^{-} = \frac{1}{r'}\, \, \, \, (\textrm{remaining part}).
\end{equation}
Also, the integral $\int dA_d$ must be taken over the area of the discharged and the remaining parts which are 
\begin{equation}
A_{d}^{+} = 4\pi r^{2} - 2\pi r\Big[(-z_{d}+r)+R\cos\theta\Big]\, \, \, \,(\textrm{discharged part}),
\end{equation}
and
\begin{equation}
A_{d}^{-} = 4\pi r'^{2} - 2\pi r'\Big[(z_{d}'+r')-R\cos\theta\Big]\, \, \, \,(\textrm{remaining part}).
\end{equation}
Thus, the bending energy of the daughter vesicle is
\begin{eqnarray}
F_{d}^{\textrm{bending}} &=& \frac{\kappa}{2}\Bigg[\Bigg(\frac{2}{r}-c_0\Bigg)^2\Big(4\pi r^{2} - 2\pi r\Big[(-z_{d}+r)+R\cos\theta\Big]\Big)\nonumber\\
&&+\Bigg(\frac{2}{r'}-c_0\Bigg)^2\Big(4\pi r'^{2} - 2\pi r'\Big[(z_{d}'+r')-R\cos\theta\Big]\Big)\Bigg].
\end{eqnarray}
While, the stretching energy is
\begin{eqnarray}
\label{stretching2}
F_{d}^{\textrm{stretching}}&=&\frac{\lambda}{2}\frac{(\Delta A_d)^{2}}{A_{d, \textrm{eq}}}\nonumber\\
&=&\frac{\lambda}{2}\frac{[(A_{d}^{+}+A_{d}^{-})-A_{d, \textrm{eq}}]^2}{A_{d, \textrm{eq}}},
\end{eqnarray}
where $A_{d, \textrm{eq}} = 4\pi r_{\textrm{eq}}^2$. Again, as shown in the SI, the free energy of the daughter vesicle can be written as a function of $r$ and $z_d$ only.

\subsection{The energy dissipation function}
 \label{subsec:dissipation function}
 
During the translocation of the daughter vesicle through the pore of its mother, the energy dissipates mostly at the pore. We can thus define the dissipation function for the daughter vesicle, in the limit of small Reynolds number, as
\begin{equation}  
\label{dissipation vesicle}
\Phi=\frac{1}{2}\zeta_{d} \dot{z_{d}}^{2} + \frac{1}{2}\zeta_{l} v_{l}^{2},
\end{equation}
where $\zeta_{d}$ is the friction coefficient of the pore on the surface of the daughter vesicle and $\zeta_{l}$ is related to the dynamic viscosity of the liquid inside the daughter vesicle. Generally, $\zeta_{d}$ and $\zeta_{l}$ should be functions of the slow variables that can be derived from the Stokesian hydrodynamics. However, in this work, we would simply treat them as constants. We have assumed that $\zeta_{d}$ is proportional to the radius of the pore $a$. Because the friction is generated at the pore's perimeter, this would allow us to write $\zeta_{d} = \alpha \cdot (2\pi a) = \alpha \cdot (2\pi R \sin\theta)$, where $\alpha$ is related to the dynamic viscosity of the lipid membranes. This assumption has been used before in our previous work.~\cite{KSIK} While, $\zeta_{l}$, which is originally caused by the liquid flow inside the daughter vesicle, is assumed to be proportional to the pore area $\pi a^{2}$. Thus, we can write $\zeta_{l} = \beta \cdot (\pi a^{2}) = \beta \cdot (\pi R^{2} \sin^{2}\theta)$, where $\beta$ has the unit of kg m$^{-2}$ s$^{-1}$. The liquid flow velocity $v_{l}$ is defined by
\begin{equation}  
\label{flow velocity}
v_{l}=\frac{\dot{V}_{\textrm{tran}}}{\pi a(t)^2}.
\end{equation}
Note that the above assumption of $\zeta_l$ and the definition of $v_l$ have been used before by two of us,~\cite{KMKD} but we should pronounce that, in this work, the pore can change its size due to the line tension energy.

 
\subsection{The daughter vesicle's equations of motion}
 \label{subsec:EOM}

In order to find the equations of motion of the daughter vesicle, we would employ the Onsager principle. The first step is to write down the Rayleighian of the system which is given by
\begin{equation}  
\mathcal{R}(\dot{r}, \dot{z_{d}}) = \Phi (\dot{r}, \dot{z_{d}}) + \dot{F}(r, z_{d}).
\end{equation}
By minimizing $\mathcal{R}$ with respect to  $\dot{r}$ and $\dot{z_{d}}$, \textit{i.e.}, $\partial\mathcal{R}/\partial\dot{r}=0$ and $\partial\mathcal{R}/\partial\dot{z_{d}}=0$, we obtain the two-coupled ordinary differential equations 
\begin{equation}  
\label{equation2_birthing}
\dot{r}=-\frac{1}{\pi a^2 \beta }\Bigg\{\Bigg[\frac{(\pi a^2)^2}{
(\partial V_{\textrm{tran}}/\partial r)^2} + \frac{a}{2(\alpha/\beta)}\Bigg(\frac{\partial V_{\textrm{tran}}/\partial z_d}{\partial V_{\textrm{tran}}/\partial r}\Bigg)^2\Bigg]
\frac{\partial F}{\partial r}-
\frac{a}{2(\alpha/\beta)}\Bigg(\frac{\partial V_{\textrm{tran}}/\partial z_d}{\partial V_{\textrm{tran}}/\partial r}\Bigg)
\frac{\partial F}{\partial z_d}\Bigg\},
\end{equation}
and
\begin{equation}  
\label{equation1_birthing}
\dot{z_{d}}=-\frac{1}{2\pi \alpha a}\Bigg[\frac{\partial F}{\partial z_{d}}-\Bigg(\frac{\partial V_{\textrm{tran}}/\partial z_d}{\partial V_{\textrm{tran}}/\partial r}\Bigg)
\frac{\partial F}{\partial r}\Bigg]
\end{equation}
that must be solved numerically.
 
\section{Results and Discussions}
 \label{sec:results and discussions}
 


\subsection{The minimum energy path for the translocation}
\label{subsec:MEP}

We first present our results of the investigation on the minimum energy path (MEP) of the translocation.
The MEP is the most probable transition pathway connecting two states of the free energy, which can be searched by the string method.~\cite{ERV1, Ren, ERV2} This method has been used in various systems, for example, to search for the MEP for the translocation of a vesicle through a pore~\cite{KMKD} and for the topological transition between two spherical vesicles and a dumbbell-shaped vesicle,~\cite{Casciola} and the application of the string method to the self-consistent field theory (SCFT) for polymers.~\cite{JTQS, STQS, SJZYT} The basic idea of the string method is that a path (string) must be evolved in compliance with the potential force in the normal directions to the path. If $\psi$ is the initial string connecting two states $a$ and $b$ on the free energy landscape $F$, the MEP can be obtained from the condition 
\begin{equation}
\label{String1}
0 = (\nabla F)^{\perp}(\psi),
\end{equation}
where $(\nabla F)^{\perp}$ is the component of $(\nabla F)$ normal to $\psi$. In order to find the MEP numerically, we first discretize the initial string connecting the two states of the translocation into a set of images where $\psi = \{\psi_i,~i = 1, 2, ..., N\}$. (Each image corresponds to different morphologies and states of the system.) Then, we evolve the images within the time step $\Delta t$ according to the dynamics 
\begin{equation}
\label{String2}
\frac{\partial \psi_i}{\partial t} = -(\nabla F)^{\perp}(\psi_i),
\end{equation}
where the free energy $F$ is an input for this dynamics. After $n$ iterations ($n$: sufficiently large), the images satisfying the differential equation Eq.~(\ref{String1}) are obtained and the length of the string is approximately constant reaching the MEP. Below we show the minimal energy pathways of the translocation obtained from the string method when the elastic moduli, the pressure inside the mother vesicle, and the size of the daughter vesicle are changed. In our model the initial radius of the daughter vesicle is denoted as $b$ and the equilibrium radius of the daughter vesicle is always set as 
$r_{\textrm{eq}} = 0.95b$. We assume that, at the initial state, the daughter vesicle is slightly discharged from the mother vesicle with $r = r_0$ and the final state corresponds to the state that the daughter is entirely expelled from the mother with $r = b$ and the pore is completely closed. The initial size of the mother vesicle at the initial state is 
$R_0$ and $R_{\textrm{eq}} = 0.98R_0$ for all presented results. The initial guess string is discretized into $N$ images where the first and last images correspond to the initial and final states, respectively.

\begin{figure}
\centering
\resizebox{1.0\textwidth}{!}{%
  \includegraphics{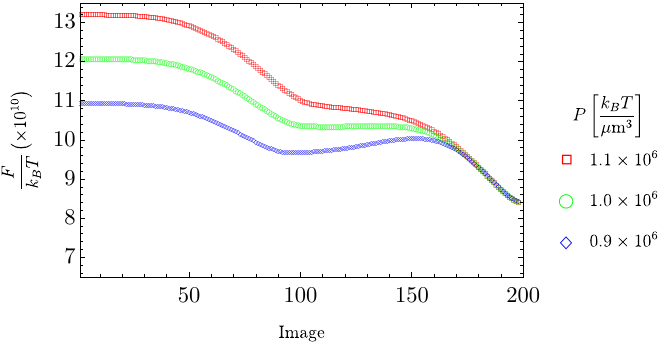}
}
\caption{Minimum energy paths of the translocation when the pressure inside the mother vesicle $P$ changes. Increasing $P$ shows the decrease in the depth of the dip where the bottom of the dip corresponds to the metastable state. The bending, stretching, and line tension moduli are fixed at  $\kappa=10\,  k_{B}T$, $\lambda=1.0\times 10^{8}\,  k_{B}T/\mu\textrm{m}^{2}$, and $\sigma=1.0\times 10^{4}\,  k_{B}T/\mu\textrm{m}$, respectively. The $\kappa/\kappa_{G}$ ratio is $-1$. The initial radius of the daughter vesicle is $b=30\, \mu\textrm{m}$. The spontaneous curvature $c_{0}$ is set as zero.}
\label{fig3}
\end{figure}

\begin{figure}
\centering
\resizebox{1.0\textwidth}{!}{%
  \includegraphics{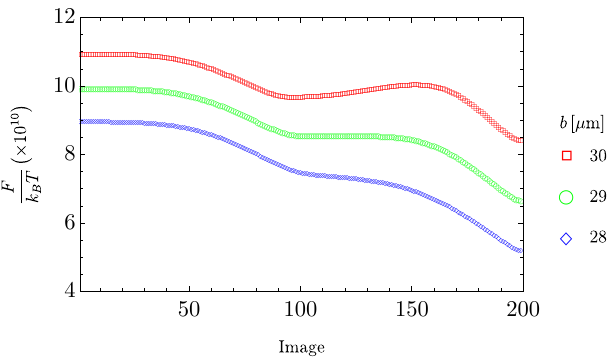}
}
\caption{Minimal energy pathways of the translocation with the variation of the radius of the daughter vesicle $b$. The daughter vesicle successfully escapes at $b = 28\, \mu\textrm{m}$. However,
increasing $b$ to $30\, \mu\textrm{m}$, the daughter is trapped at the pore. We have fixed the bending, stretching, and line tension moduli as $\kappa=10\,  k_{B}T$, $\lambda=1.0\times 10^{8}\,  k_{B}T/\mu\textrm{m}^{2}$, and $\sigma=1.0\times 10^{4}\,  k_{B}T/\mu\textrm{m}$, respectively. $P=0.9\times 10^{6}\,  k_{B}T/\mu\textrm{m}^{3}$, $\kappa/\kappa_{G} = -1$, and $c_{0} = 0$.}
\label{fig4}
\end{figure}

We first investigate the effects of the pressure inside the mother vesicle $P$ and the size of the daughter vesicle $b$ on the translocation process as shown in Figs.~\ref{fig3} and~\ref{fig4}. In order to see the effects, we have fixed the bending, stretching, and line tension moduli as  $\kappa=10\,  k_{B}T$, $\kappa_G = -10\,  k_{B}T$, $\lambda=1.0\times 10^{8}\,  k_{B}T/\mu\textrm{m}^{2}$, and $\sigma=1.0\times 10^{4}\,  k_{B}T/\mu\textrm{m}$, respectively. The initial string is discretized into $N = 200$ images. At the initial state, $R_{0}=40\, \mu\textrm{m}$ and  $r_{0}=0.1\, \mu\textrm{m}$. The minimum energy paths obtained from the string method when $b$ is fixed at $30\, \mu\textrm{m}$ show clearly that the depth of the dip where the daughter vesicle is trapped in the basin of the metastable state decreases as $P$ increases. The plot shows that the daughter vesicle successfully escapes from its mother at $P=1.1\times 10^{6}\,  k_{B}T/\mu\textrm{m}^{3}$ as shown in 
Fig.~\ref{fig3}. For fixed $P=0.9\times 10^{6}\,  k_{B}T/\mu\textrm{m}^{3}$, the daughter vesicle with $b=28\, \mu\textrm{m}$ is successfully discharged. Further increasing $b$ to $30\, \mu\textrm{m}$, the daughter is stuck in the basin of the metastable state. This effect can be seen in the minimal energy pathways shown in Fig.~\ref{fig4}.

\begin{figure}
\centering
\resizebox{1.0\textwidth}{!}{%
  \includegraphics{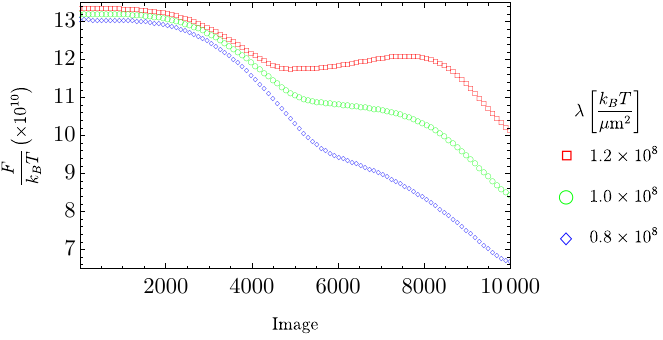}
}
\caption{MEPs of the translocation of the daughter vesicle when the stretching modulus $\lambda$ changes. The bending and line tension moduli are fixed at $\kappa = 10\,  k_{B}T$ and $\sigma=1.0\times 10^{4}\,  k_{B}T/\mu\textrm{m}$, respectively. $b = 30\, \mu\textrm{m}$, and $P=1.1\times 10^{6}\,  k_{B}T/\mu\textrm{m}^{3}$.
The $\kappa/\kappa_{G}$ ratio is $-1$. The spontaneous curvature $c_{0}$ is set as zero. The daughter can successfully escape at $\lambda=1.0\times 10^{8}$ and $0.8\times 10^{8}\,  k_{B}T/\mu\textrm{m}^{2}$.}
\label{fig5}
\end{figure}

\begin{figure}
\centering
\resizebox{1.0\textwidth}{!}{%
  \includegraphics{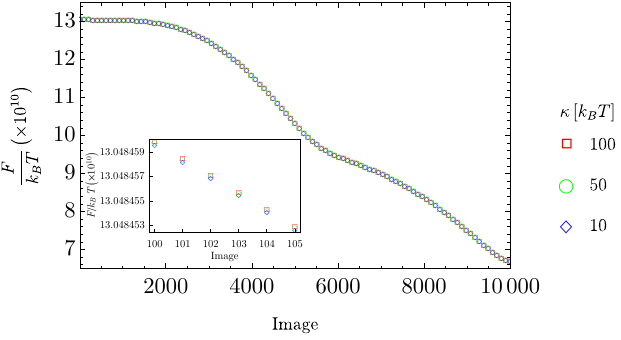}
}
\caption{Minimum energy paths with the variation of $\kappa$. $\lambda$ and $\sigma$ are $0.8\times 10^{8}
\,  k_{B}T/\mu\textrm{m}^{2}$ and
$1.0\times 10^{4}\,  k_{B}T/\mu\textrm{m}$, respectively. $P$ is fixed at $1.1\times 10^{6}\,  k_{B}T/\mu\textrm{m}^{3}$. All paths show the successful translocation of the daughter vesicle but the variation among the paths cannot be seen on the large scale. (Inset) The variation of the paths is shown at the small scale.}
\label{fig6}
\end{figure}

\begin{figure}
\centering
\resizebox{1.0\textwidth}{!}{%
  \includegraphics{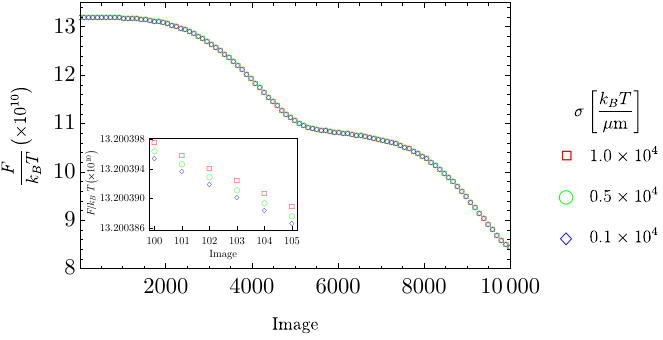}
}
\caption{MEPs of the translocation process with changes in $\sigma$, while $\kappa$ and $\lambda$ are fixed at $10\,  k_{B}T$ and $1.0\times 10^{8}
\,  k_{B}T/\mu\textrm{m}^{2}$, respectively. $P = 1.1\times 10^{6}\,  k_{B}T/\mu\textrm{m}^{3}$. All curves show that the daughter vesicle can pass through the mother vesicle but there is no significant variation among those pathways. (Inset) The difference of the paths can be seen at the small scale.}
\label{fig7}
\end{figure}

The dependences of the translocation of the daughter vesicle on the bending, stretching, and line tension moduli are shown in Figs.~\ref{fig5},~\ref{fig6}, and~\ref{fig7}. All plots are evaluated at $R_0 = 40\, \mu\textrm{m}$, $b = 30\, \mu\textrm{m}$, $P=1.1\times 10^{6}\,  k_{B}T/\mu\textrm{m}^{3}$, and $N = 10,000$ images. Figure~\ref{fig5} shows the MEPs for the various values of the stretching modulus $\lambda$ when the bending and line tension moduli are fixed at  $\kappa=10\,  k_{B}T$, and $\sigma=1.0\times 10^{4}\,  k_{B}T/\mu\textrm{m}$, respectively. As $\lambda=1.2\times 10^{8}\,  k_{B}T/\mu\textrm{m}^{2}$ the MEP shows that the daughter vesicle is trapped but it can escape at $\lambda=1.0\times 10^{8}$ and $0.8\times 10^{8}\,  k_{B}T/\mu\textrm{m}^{2}$. Figure~\ref{fig6} presents the MEPs for $\kappa = 10, 50, 100\,  k_{B}T$. $\lambda$ and $\sigma$ are fixed at $0.8\times 10^{8}\,  k_{B}T/\mu\textrm{m}^{2}$ and $1.0\times 10^{4}\,  k_{B}T/\mu\textrm{m}$, respectively. The MEPs show that the daughter vesicle successfully passes through the pore for all selected $\kappa$ but the variations of $\kappa$ did not change the pathway remarkably. The effect of changing $\kappa$ can be seen only at the small scale as shown in the inset of Fig.~\ref{fig6}. The similar behavior has been observed for the changes in $\sigma$ while $\kappa=10\,  k_{B}T$ and $\lambda = 1.0\times 10^{8}\,  k_{B}T/\mu\textrm{m}^{2}$ as shown in Fig.~\ref{fig7}. This means that the bending and line tension energies did not play a significant role in the translocation of the daughter vesicle. As the results published in our previous work,~\cite{KSIK} the line tension energy also affects the process drastically but the value of $\sigma$ we assumed is overestimated, i.e., around $10^{8}\,  k_{B}T/\mu\textrm{m}$, while, in this work, $\sigma$ is selected from the experimental reports. 



\subsection{Phase Diagram}
\label{subsec:phase diagram}

\begin{figure}
\centering
\resizebox{1.0\textwidth}{!}{%
  \includegraphics{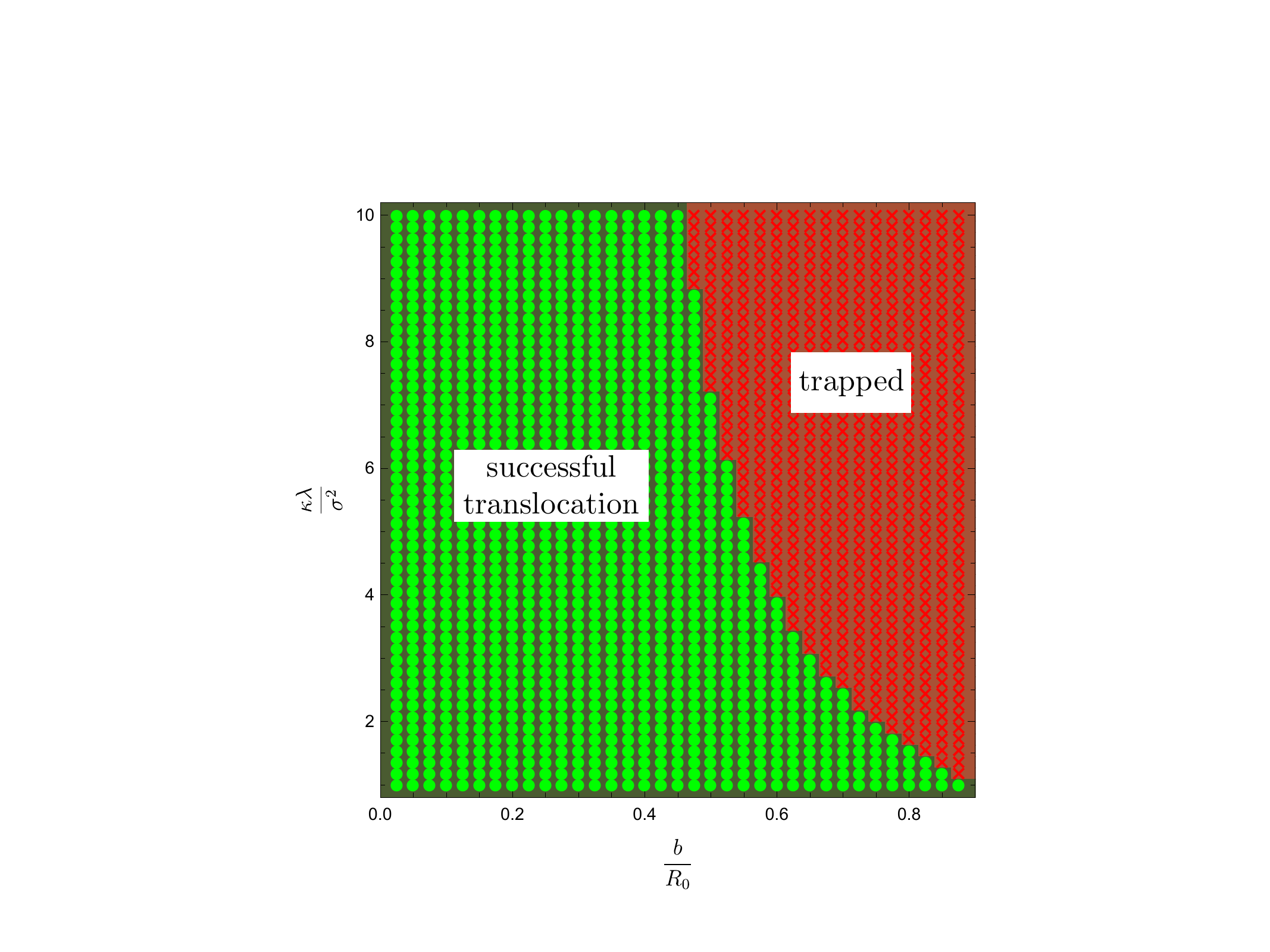}
}
\caption{Phase diagram of the translocation of the daughter vesicle at the dimensionless moduli $\kappa \lambda/\sigma^2$ and $b/R_0$. $R_0 = 40\, \mu\textrm{m}$, $P=2.0\times 10^{5}\,  k_{B}T/\mu\textrm{m}^{3}$, $\kappa=10\,  k_{B}T$, and $\sigma=1.0\times 10^{4}\,  k_{B}T/\mu\textrm{m}$. $b$ is changed from $1\, \mu\textrm{m}$ to $35\, \mu\textrm{m}$. $\lambda$ varies from $1.0\times 10^{7}\,  k_{B}T/\mu\textrm{m}^{2}$ to $1.0\times 10^{8}\,  k_{B}T/\mu\textrm{m}^{2}$.}
\label{fig8}
\end{figure}

Figure~\ref{fig8} shows the phase diagram of the translocation process determined from the string method at the dimensionless elastic moduli $\kappa \lambda/\sigma^2$ and $b/R_0$. The system parameters are $R_0 = 40\, \mu\textrm{m}$, $P=2.0\times 10^{5}\,  k_{B}T/\mu\textrm{m}^{3}$, $\kappa=10\,  k_{B}T$, and $\sigma=1.0\times 10^{4}\,  k_{B}T/\mu\textrm{m}$. $b$ is varied from $1\, \mu\textrm{m}$ to $35\, \mu\textrm{m}$ and $\lambda$ is changed from $1.0\times 10^{7}\,  k_{B}T/\mu\textrm{m}^{2}$ to $1.0\times 10^{8}\,  k_{B}T/\mu\textrm{m}^{2}$. The phase diagram shows the region of the successful translocation of the daughter vesicle (indicated with filled circle, green color) separated by the region that the daughter vesicle is trapped at the pore (indicated with cross, red color). The phase diagram seems to show that even the size of the daughter vesicle is quite large, the daughter vesicle can still pass through the mother vesicle if the pressure inside the mother vesicle is sufficiently high. However, when we increase the stretching modulus, the depth of the basin of the metastable state gets deeper (as we have shown in Fig.~\ref{fig5}). This causes the decrease in the successful translocation region. This result significantly differs from the phase diagram in our previous work~\cite{KSIK} that shows an abrupt change in the energy barrier due to the increased line tension energy. While, in this improved model, we have shown that the line tension energy did not play an important role (Fig.~\ref{fig7}).

\subsection{Translocation Time}
\label{subsec:time}

\begin{figure}
\centering
\resizebox{1.0\textwidth}{!}{%
  \includegraphics{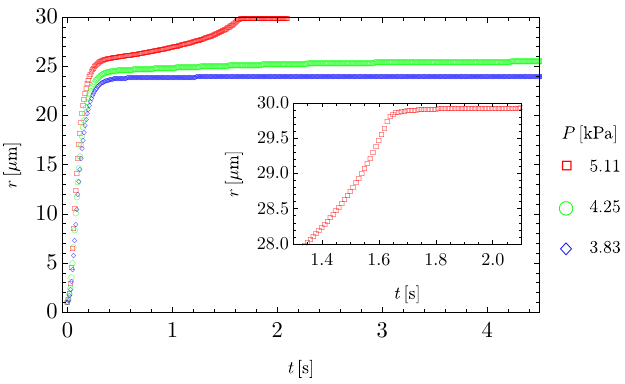}
}
\caption{The motion of the daughter vesicle with different values of the pressure inside the mother vesicle. The plots are evaluated at $R_0 = 40\, \mu\textrm{m}$, $b = 30\, \mu\textrm{m}$. The elastic moduli are selected as $\kappa = 10\, k_B T$, $\kappa/\kappa_G = -1$, $\lambda = 1.0 \times 10^8\, k_BT/\mu\textrm{m}^2$, and $\sigma = 1.0 \times 10^4\, k_BT/\mu\textrm{m}$. $\alpha = 10^3$ kg m$^{-1}$ s$^{-1}$ and $\alpha/\beta = 5 \times 10^{-5}\, \textrm{m}$. The plots are evaluated at $T = 35^{\circ} \textrm{C}$ and $c_0 = 0$. (Inset) The daughter vesicle gradually expands the translocation part from $t \approx 1.65\, \textrm{s}$ until it can detach at $t \approx 2.1\, \textrm{s}$.}
\label{fig9}
\end{figure}
\begin{figure}
\centering
\resizebox{1.0\textwidth}{!}{%
  \includegraphics{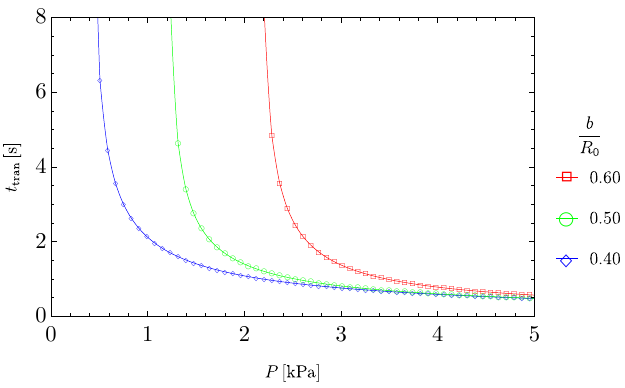}
}
\caption{The dependence of the translocation time on $P$ when $b/R_0$ changes. $\kappa = 10\, k_B T$, $\kappa/\kappa_G = -1$, $\lambda = 1.0 \times 10^8\, k_BT/\mu\textrm{m}^2$, and $\sigma = 1.0 \times 10^4\, k_BT/\mu\textrm{m}$. $\alpha$ is selected as $10^3$ kg m$^{-1}$ s$^{-1}$ and $\alpha/\beta = 5 \times 10^{-5}\, \textrm{m}$. $T = 35^{\circ} \textrm{C}$ and $c_0 = 0$.}
\label{fig10}
\end{figure}

In this section, we investigate the translocation time of the daughter vesicle. The system parameters are $R_0 = 40\, \mu\textrm{m}$, $R_{\textrm{eq}} = 0.98R_0$, and $c_0 = 0$. The elastic moduli are selected as $\kappa = 10\, k_B T$, $\kappa/\kappa_G = -1$, $\lambda = 1.0 \times 10^8\, k_BT/\mu\textrm{m}^2$, and $\sigma = 1.0 \times 10^4\, k_BT/\mu\textrm{m}$. The value of $\alpha$ for a spherically rigid daughter vesicle can be estimated from our formula derived in the Appendix of the previous work,~\cite{KSIK} which says 
\begin{equation}
    \alpha = \frac{\lambda t_\textrm{tran}}{2c}\frac{\Delta A}{A_{\textrm{eq}}},
    \end{equation}
where $c$ is the radius of the rigid vesicle and $\Delta A$ is the excess area. One of us (Sakuma) has measured the dependence of $t_\textrm{tran}$ on the size of the daughter vesicle in the model system of self-reproducing DLPE/DPPC vesicle. For the case where the radii of the mother and daughter vesicles are about $14$ and $6.5\, \mu\textrm{m}$, respectively, $t_\textrm{tran}$ is around $1\, \textrm{s}$.~\cite{YS} Varying sizes of the two vesicles causes a change in  $t_\textrm{tran}$. The maximum $t_\textrm{tran}$ found in the evaluations is about 20 s. Regarding these measures, the calculated $\alpha$ is of the order of $10^3 - 10^4$ kg m$^{-1}$ s$^{-1}$. In order to evaluate $t_\textrm{tran}$ for the deformed daughter vesicle, we have simply selected $\alpha$ as $10^3$ kg m$^{-1}$ s$^{-1}$ and $\alpha/\beta = 5 \times 10^{-5}\, \textrm{m}$ is used. All plots will be evaluated at $T = 35^{\circ} \textrm{C}$. Figure~\ref{fig9} shows the plot of $t$ vs $r$ for various values of $P$ at $b = 30\, \mu\textrm{m}$. Clearly, at $P = 3.83\, \textrm{kPa}$ and $4.25\, \textrm{kPa}$, the daughter vesicle cannot evacuate from its mother. Further increasing $P$ to $5.11\, \textrm{kPa}$, the daughter can pass through the mother with the translocation time $t_{\textrm{tran}}\approx 2.1\, \textrm{s}$. 
Investigating the path at $P = 5.11\, \textrm{kPa}$, we might divide the translocation pathway into three stages. From $t = 0$ to $t\approx 0.3\, \textrm{s}$, the inflation rate $(dr/dt)$ of the daughter vesicle is extremely high because in the early stage of the translocation $V_{\textrm{tran}}$ is much smaller than $V_0$. The pressure contribution is very high causing the rapid pushing out the daughter vesicle. Later, the rate does not change so much until $t\approx 1.65\, \textrm{s}$. While, at the last stage of the translocation, because the pressure contribution becomes less and less, the daughter slowly grows its size until it can be detached from the mother. Figure~\ref{fig10} shows the dependence of the $t_{\textrm{tran}}$ on $P$ with the variation of $b/R_0$. Clearly, at fixed $P$, increasing $b/R_0$ requires the longer time for the daughter vesicle to escape. Our simulations suggest the critical pressure for the successful translocation as $P_c \approx 2.086, 1.128, 0.358\, \textrm{kPa}$ for $b/R_0 = 0.60, 0.50, 0.40$, respectively.

\section{Conclusions}
 \label{sec:conclusions}
 
In this work, we have developed a model for the translocation of a daughter vesicle in a model system of self-reproducing vesicles first proposed in our previous work. The main and significant modification is to consider the daughter vesicle as a deformable object instead of a rigid body. We thus derive the free energy of the system constituting of the elastic moduli within the Helfrich formalism where the pressure inside the mother vesicle is a driving force causing the evacuation of the daughter vesicle trough a pore. By using the string method, the present model suggests the successful translocation in which the elastic moduli are in a good agreement with the available experimental data. Clearly, our improved model eliminates the discrepancy appeared in the previous work. The model also suggests that the stretching energy plays a significant role in the translocation. This result is inconsistent with what we reported in the previous work which shows the competition between the stretching and line tension energies. However, we should pronounce that the value of the line tension modulus $\sigma$ in the previous work is overestimated, i.e, it is of the order of $10^8\, k_BT/\mu\textrm{m}$, while the experimental reports show that $\sigma$ is of the order of  $10^3 - 10^4\, k_BT/\mu\textrm{m}$. Then, we believe that our latest improvement reflects more realistic phenomenon. In order to study the kinetics of the system, we have employed the Onsager principle. The kinetic equations of the daughter vesicle are derived. We found that, by selecting the friction coefficients related to the friction against the moving daughter vesicle due to the inner edge of the pore and liquid flows inside the daughter vesicle, the translocation time can be reduced when we increase the pressure inside the mother vesicle, or the initial size of the daughter vesicle decreases.

\section*{Supplementary Material}

We have described the methods for eliminating the degrees of freedom of the system in the supplementary material.

 \begin{acknowledgments}
M.K. and P.K. acknowledge the financial support from Ramkhamhaeng University through Research Institute.
 
\end{acknowledgments}

\section*{data availability}

The data that support the findings of this study are available from the corresponding author upon reasonable request.

\appendix

\section{Derivation of the $\kappa_{G}/\kappa$ ratio}
 \label{ratio}

Let us consider an orthogonal coordinate system at a point on a curved surface. The two geodesic curvatures are denoted as $c_1$ and $c_2$. The bending elastic energy density at this point can be described as 
\begin{equation}
    \label{density}
    h(c_1, c_2) = \frac{1}{2}\begin{pmatrix} c_{1}&c_{2}\end{pmatrix}    
    \begin{pmatrix} a_{11}&a_{12}\\a_{21}&a_{22} \end{pmatrix}\begin{pmatrix} c_{1}\\c_{2} \end{pmatrix} \equiv \frac{1}{2}\mathbf{c}^\mathrm{T}\mathbf{A}\mathbf{c},
\end{equation}
where the linear term is neglected because of the vanishing spontaneous curvature of symmetric membrane.

We assume that the membrane is a liquid membrane and isotropic inside the membrane surface. Then, the coefficient matrix $\mathbf{A}$ can be expressed as
\begin{equation}
    \label{isotropic}
    \mathbf{A} = \begin{pmatrix} a&0\\0&a \end{pmatrix}.
\end{equation}

Substituting Eq.~\ref{isotropic} into Eq.~\ref{density}, we obtain
\begin{eqnarray}
    \label{h1}
    h(c_1, c_2) &=& \frac{a}{2}(c_1^{2} + c_2^{2})\nonumber\\
    & = & 2aH^2 - aK,
\end{eqnarray}
where we have defined the mean and Gaussian curvatures as
\begin{equation}
    H = \frac{1}{2}(c_1 + c_2),
\end{equation}
and
\begin{equation}
    K = c_{1}c_{2},
\end{equation}
respectively. Comparing Eq.~\ref{h1} with the Helfrich’s bending elastic model
\begin{equation}
    h_{\textrm{Helfrich}} = 2\kappa H^2 + \kappa_{G}K.
\end{equation}
We thus obtain $\kappa = -\kappa_{G} = a$. Finally, we can conclude that
\begin{equation}
    \kappa_{G}/\kappa = -1.
\end{equation}


\begin{thebibliography}{00}
\bibitem{YS} Y.~Sakuma, The measured diameters of the mother and daughter vesicles are 28.2$\, \mu\textrm{m}$ and 13.1$\, \mu\textrm{m}$, respectively in the model system for self-reproducing DLPE/DPPC vesicle showing the successful translocation of the daughter vesicle with translocation time around a second. The translocation time can be up to around 20 s for larger daughter vesicle.
\end{thebibliography}

\end{document}